\documentclass[%
 reprint,
superscriptaddress,
 amsmath,amssymb,
pra,
]{revtex4-2}

\usepackage[utf8]{inputenc}
\usepackage{graphicx}
\usepackage{amsmath}
\usepackage{amsfonts}
\usepackage{amsthm}
\usepackage{subcaption}
\usepackage{enumitem}
\usepackage{braket}
\usepackage{color}
\usepackage{hyperref}
\usepackage{fullpage}
\usepackage[title]{appendix}

\newtheorem*{thm*}{Theorem}

\newtheorem{lem}{Lemma}

\theoremstyle{definition}

\theoremstyle{remark}

\begin{document}

\title{Quantum advantage for differential equation analysis}

\author{Bobak Toussi Kiani}
\affiliation{Department of Electrical Engineering and Computer Science, Massachusetts Institute of Technology, Cambridge, MA 02139}
\affiliation{Research Laboratory of Electronics, Massachusetts Institute of Technology, Cambridge, MA 02139}
\author{Giacomo De Palma} 
\affiliation{Research Laboratory of Electronics, Massachusetts Institute of Technology, Cambridge, MA 02139}
\affiliation{Scuola Normale Superiore, 56126 Pisa, Italy}
\author{Dirk Englund}
\affiliation{Department of Electrical Engineering and Computer Science, Massachusetts Institute of Technology, Cambridge, MA 02139}
\affiliation{Research Laboratory of Electronics, Massachusetts Institute of Technology, Cambridge, MA 02139}
\affiliation{QuEra Computing, Boston, MA 02143}
\author{William Kaminsky}
\affiliation{Independent Contractor, Cambridge, MA 02139}
\author{Milad Marvian}
\affiliation{Research Laboratory of Electronics, Massachusetts Institute of Technology, Cambridge, MA 02139}
\affiliation{Department of Electrical and Computer Engineering, Center for Quantum Information and Control (CQuIC), University of New Mexico,  Albuquerque, NM 87131}
\author{Seth Lloyd}
\affiliation{Department of Mechanical Engineering, Massachusetts Institute of Technology, Cambridge, MA 02139}
\affiliation{Research Laboratory of Electronics, Massachusetts Institute of Technology, Cambridge, MA 02139}

\date{\today}

\begin{abstract}
Quantum algorithms for differential equation solving, data processing, and machine learning potentially offer an exponential speedup over all known classical algorithms.  However, there also exist obstacles to obtaining this potential speedup in useful problem instances.   The essential obstacle for quantum differential equation solving is that outputting useful information may require difficult post-processing, and the essential obstacle for quantum data processing and machine learning is that inputting the data is a difficult task just by itself. In this study, we demonstrate, when combined, these difficulties solve one another.  We show how the output of quantum differential equation solving can serve as the input for quantum data processing and machine learning, allowing dynamical analysis in terms of principal components, power spectra, and wavelet decompositions.  To illustrate this, we consider continuous time Markov processes on epidemiological and social networks.  These quantum algorithms provide an exponential advantage over existing classical Monte Carlo methods.
\end{abstract}
\maketitle

\section{Introduction}

One of the primary proposed applications of quantum computers is the solution of linear differential equations on high-dimensional spaces.    The ability of quantum computers to represent $N$-dimensional vectors as the state of $\log_2 N$ qubits, and to perform linear algebraic transformations of those states in time $\mathrm{poly}( \log N)$, then translates into a potential exponential speedup over classical algorithms for solving such high-dimensional differential equations.  The output of the quantum computer presents the solution to the equation as a quantum `history state' that is a superposition of the solution at different points in time.   The problem then arises: How do we extract useful information from that quantum solution state?    We can measure the expectation value of different quantities of interest: however, in the example of Markov chains,
such expectation values can often be evaluated efficiently classically via Monte Carlo techniques.  To obtain a quantum advantage that reveals essential features of the solution to the linear differential equations, we need to perform quantum post-processing on the history state.
 
Quantum machine learning and data processing algorithms provide potential exponential speedups over classical counterparts for methods such as high-dimensional regression, principal component analysis, and support vector machines \cite{biamonte2017quantum,lloyd2014quantum,schuld2016prediction,rebentrost2014quantum}.   A basic problem with such quantum algorithms is that the input to the algorithm is a quantum state that encodes the classical data, and to construct such a state requires the implementation of a large-scale quantum random access memory (qRAM), a difficult technological task. The central observation of this study is that the problem with quantum linear equation solvers -- they give quantum states as output -- and the problem with quantum machine learning and data processing algorithms -- they require quantum states as input -- effectively solve each other: the output from the quantum linear equation solver can be used as the input to the quantum machine learning or data processing algorithm.  In particular, we show that the history-state quantum solution to high-dimensional linear differential equations takes exactly the form needed to perform quantum analysis of the solution via quantum machine learning and data analysis.    We show how to produce the singular values and singular vectors of the solution via quantum principal component analysis, and how to extract the power spectrum of the solution by performing quantum Fourier transforms.   This analysis reveals the dominant components of the time evolution, corresponding to large singular values and eigenvalues of the transition matrix with small negative real part.  Finally we show how to perform quantum wavelet analysis to reveal rapid transitions and emergent features in the solution at different time scales. These quantum algorithms for post-processing the solution of linear differential equations can yield an exponential speedup over classical methods, and could potentially be performed on near term intermediate scale quantum computers. 

The quantum post-processing of the history state to reveal salient features of the history can be applied to any linear differential equation.  To show how quantum post-processing reveals such features, we focus on the case of continuous time Markov chains on high-dimensional spaces, with an emphasis on the spread of disease and opinion in complex social networks.  
Previous quantum algorithms for Monte Carlo methods yielded a square root speedup over classical algorithms \cite{montanaro2015quantum}.   By contrast, the quantum algorithms presented here for analyzing linear dynamics in terms of singular values, power spectra, and wavelets represent an exponential speedup over existing classical Monte Carlo methods.

\section{Results}

\subsection{Quantum algorithms for linear differential equations}

Quantum numerical algorithms solve differential equations by quantizing classical numerical procedures and performing matrix operations on finite, high-dimensional state spaces \cite{berry2014high,berry2017quantum,lloyd2020quantumnonlinear,liu2020efficient}. We write a general linear differential equation for a vector in $\mathbb{R}^N$ with $N\gg1$ in the form 

\begin{equation}
    \frac{d \vec x(t)}{dt} = {\cal M} \vec x(t) + \vec c
    \label{eq:generalDiff}
\end{equation}
where $x(t)$ represents the state of the system at time $t$, ${\cal M}$ is the $N\times N$ differential equation matrix, and $c$ is a forcing term. For example, in the case of Markov models, the state space is represented by a probability vector $x(t)$ with entries $x_i(t)$ indicating the probability of the system existing in state $i$ at time $t$, and ${\cal M}$ the transition matrix. 

Well-known quantum algorithms \cite{berry2014high,berry2017quantum}
can solve linear differential equations of the form 
of Eq. \ref{eq:generalDiff}, where  $ {\cal M}$ is a sparse matrix, by applying the quantum algorithm for linear systems of equations \cite{harrow2009quantum,childs2015quantum}. The algorithms of Refs. \cite{berry2014high,berry2017quantum} are based on a simple underlying idea, but require highly nontrivial technical improvements in order to achieve a low computational complexity.
For the sake of a clearer exposition, we present here only the basic idea, and refer the reader to Refs. \cite{berry2014high,berry2017quantum} for a complete presentation of the algorithms.
We consider the differential equation (Eq. \ref{eq:generalDiff}) for $0\le t \le t_{\max}$.
The idea is based on the standard classical 
methods for discretizing Eq. \ref{eq:generalDiff} in $T$ time steps each of size $h = t_{\max}/T$ and re-framing it as the solution to an equation of the form

\begin{equation}
    A \vec x = \vec b
    \label{eq:linearInverse}
\end{equation}{}
where 
\begin{equation}
  \vec x = \begin{pmatrix} 
              \vec x_0  \\ 
              \vec x_1  \\ 
              \vec x_2  \\ 
              \vdots \\ 
              \vec x_T 
  \end{pmatrix} = 
  \begin{pmatrix}
              \vec x(0) \\ 
              \vec x(h) \\ 
              \vec x(2h) \\ 
              \vdots \\ 
              \vec x(Th)
  \end{pmatrix}
  \label{eq:vecSolution}
\end{equation}
is a `vector of vectors' that contains the values of the
state vectors for the system,
$\vec x_\ell = \vec x(t_\ell)$, 
at different moments of discretized time $t_\ell = \ell h$. $A$ is a matrix that represents the updating action of 
the discretized differential operator. The form of $A$ and $\vec b$ depends on which discretization method one employs
for the differential equation (\textit{e.g.,} Euler forward,
Euler backward, Crank-Nicolson, etc.).   The simplest method is
Euler forward, where 
\begin{equation}
  A = \begin{pmatrix} 
              I &0 &\ldots  & 0 & 0 \\ 
               -( I + {\cal M} \Delta t) & I & \ldots  & 0 & 0\\ 
               & & \ddots & & & \\
              0 & 0 &\ldots & I & 0 \\
              0 & 0 &\ldots & -( I + {\cal M} \Delta t) & I
  \end{pmatrix}
  \label{eq:quantumNumericalMatrix}
\end{equation}
and
\begin{equation}
  \vec b =  
  \begin{pmatrix} 
              \vec x_0 \\ 
              \vec c \\ 
              \vdots \\ 
              \vec c
  \end{pmatrix}
\end{equation}
The solution to the differential equation is then obtained by inverting
the matrix in Eq. \ref{eq:linearInverse}:
\begin{equation}
    \vec x = A^{-1} \vec b
\end{equation}
Roughly speaking, the quantum algorithms of \cite{berry2014high,berry2017quantum} map the classical states onto quantum states,
$\vec x  \rightarrow |x\rangle$, $\vec b \rightarrow |b\rangle$, and
solve the problem using quantum matrix inversion to construct
the normalized version of the unnormalized quantum
state $|x\rangle = A^{-1} |b\rangle$, which
represents a quantum superposition of the solutions of 
Eq. \ref{eq:generalDiff} at different points in time.

We will base our results on the algorithm of Ref. \cite{berry2017quantum}, which has the lowest computational complexity.
This algorithm sets $T=\left\lceil t_{\max}\left\|\mathcal{M}\right\|\right\rceil$, employs the Taylor expansion of the matrix exponential to solve the differential Eq. \ref{eq:generalDiff}, and produces a coherent superposition of the quantum state $|x\rangle$ with some garbage state associated to the terms of the Taylor series.
We show in Appendix \ref{app:runtime_proofs} that the normalized version $|\bar{x}\rangle = |x\rangle/\||x\rangle\|$ of $|x\rangle$ can be recovered from the quantum state produced by the algorithm of Ref. \cite{berry2017quantum} with $O(1)$ success probability.

\paragraph*{\textbf{Form of the quantum solution}}
The (unnormalized) quantum state $\ket{x}$ contains registers for the states and time-steps:
\begin{equation}
    \ket{x} = \sum_{i=0}^T \ket{x_i} \ket{i}
    \label{eq:quantumDataMatrix}
\end{equation}
where the entries of $\ket{x_i}$ are the values of the state for timestep $\ket{i}$. $\ket{x}$ is a quantum `history state': a superposition of the different timesteps $\ket{i}$,
correlated/entangled with the corresponding state vectors $\ket{x_i}$ at that timestep.


The quantum state in Eq. \ref{eq:quantumDataMatrix} can now be post-processed using efficient quantum algorithms such as those for wavelet transforms or quantum machine learning. 
Alternatively, history states encoding the evolution of an arbitrary quantum circuit at different times can be created by preparing the ground state of a local Hamiltonian \cite{feynman1985quantum,kitaev2002classical} and then subsequently post-processed. 

\paragraph*{\textbf{Computational runtime and scaling of the error}}

The computational complexity of the quantum algorithm of Ref. \cite{berry2017quantum} is linear in the condition number and the time,
and logarithmic in the error.
We show in Appendix \ref{app:runtime_proofs} that, by running this algorithm and by suitably projecting the generated quantum state, we can obtain a quantum
state that is $\epsilon$-close to the state $|x\rangle$ in $2$-norm with
\begin{equation}
O\left(\kappa' \,\mathrm{poly}\log\left(\left(1+\frac{t_{\max}\left\|\vec{c}\right\|}{\left\|\vec{x}(0)\right\|}\right)\frac{\kappa'\,N}{\epsilon}\right)\right)
\end{equation}
elementary quantum gates.
Here $\kappa' = \kappa\,t_{\max}\left\|\mathcal{M}\right\|s$, $\kappa$ is the condition number of the matrix that is used to
diagonalize ${\cal M}$ and $s$ is the sparsity of ${\cal M}$. 

Furthermore, input states $\ket{b}$ for differential equation solvers can often be efficiently constructed via efficient algorithms that apply sparse matrix or local operations. This is especially true in the case of Markov chains, further discussed later, where initial states are often supported on a sparse number of entries or locally uniformly throughout the possible set of states (see Appendix \ref{app:input_preparation}).

\paragraph*{\textbf{Application to the Schr\"{o}dinger equation}}
In the case where ${\cal M}$ is anti-Hermitian, then Eq. \ref{eq:generalDiff} is the Schr\"{o}dinger equation \cite{feynman1985quantum}. In this case, the Fourier analysis of the history state reveals the eigenvalues and eigenvectors of ${\cal M}$. For example, when ${\cal M}$ is the Feynman Hamiltonian, the history state encodes the history of quantum computation and the Fourier analysis reveals its eigenstates. Finding the eigenstates of the Feynman Hamiltonian is at least as hard as performing the quantum computation \cite{lloyd2008robustness}.

\subsection{Continuous time Markov chains}

To illustrate the power of our methodology, in this study, we focus on differential equations for continuous time Markov chains which update probability distributions over state space by assuming that the probability of making a transition to the next state only depends on the current state of the Markov chain. 
Here, we focus on Markov chains that represent dynamics of complex networks, particularly those modeling the spread of opinion and disease. The dynamics of complex networks are modeled via Markovian techniques by assuming that each node in a network exists in one of $q$ states. A central challenge in this setup is in handling the dimension of the Markov state which grows exponentially with the number of nodes in a graph, often rendering the problem intractable for classical computers. For example, in epidemiology, modeling the dynamics of infection and recovery for a system of individuals whose interactions make up a complex network of $n$ nodes is a hard computational problem that involves predicting the behavior of a very large $q^n$ dimensional continuous time probabilistic dynamics \cite{pastor2015epidemic,sahneh2013generalized}. The exact solution for the dynamics of such epidemiological models lies beyond the realm of capability of even the most powerful classical computers.  Consequently, classical approaches typically rely on various approximations, such as mean field theory \cite{pastor2015epidemic}.

Continuous time Markov chains are defined by differential equations of the form
\begin{equation}
    \frac{d \vec x(t)}{dt} = {\cal M} \vec x(t)
    \label{eq:CTMC_diff_eq}
\end{equation} 
where $\vec x(t)$ is the vector of probabilities for the underlying state
of the system at time $t$, and ${\cal M}$ is a matrix of transition rates \cite{anderson2012continuous,kolmogoroff1931analytischen}.

Eq. \ref{eq:CTMC_diff_eq} is precisely of the form required in Eq. \ref{eq:generalDiff} for efficiently solving differential equations using quantum algorithms.
As shown in Appendix \ref{app:runtime_proofs}, the quantum algorithm of Ref. \cite{berry2017quantum} can be employed to produce a normalized version of the quantum state $\ket{x} = \sum_{i=0}^T \ket{x_i} \ket{i}$ storing the state probabilities at discretized times.   Note that the quantum algorithm represents the vector of probabilities as quantum vector of probability amplitudes. 

\paragraph*{\textbf{Generalization to non-Markovian models}}
We can generalize our algorithms to incorporate prior histories in a ``non-Markovian'' model.
The Markovian nature to the methods for solving differential equations
is reflected in the form of the matrix $A$ in Eq. \ref{eq:quantumNumericalMatrix} above.
The fact that the Euler forward method for discretizing a differential
equation depends only on the current and previous state of the system
implies that $A$ only has entries on the diagonal and directly below the
diagonal.   If we wish to include influences on the present from further
in the past (including the distant past), then we can simply add additional
entries to each row: adding a matrix entry $A_{ij}$, $j < i$ to the
$i$'th row of $A$ allows the state of the system at time $j<i$ to influence
the updating at time $i$.
This change cannot be directly incorporated in the algorithm of Ref. \cite{berry2017quantum}, which relies on the Taylor expansion of the matrix exponential, but it can easily be incorporated in the previous algorithm of Ref. \cite{berry2014high}, which directly solves Eq. \ref{eq:linearInverse}.

\subsection{Quantum post-processing}
The solution of our quantum differential equation solver $\ket{x}$ exists in a very high dimensional Hilbert space, and here, we discuss methods to obtain useful information from $\ket{x}$ using various quantum algorithms that can offer exponential speedups over classical counterparts. In this study, we focus on the particular case of post-processing outputs from continuous time Markov chain models. The algorithms we list here are by no means comprehensive of the full catalog of algorithms available to quantum computer scientists for extracting information from these states. 

\paragraph*{\textbf{Post-processing: expectation values of quantities}}
The most basic information that can be extracted from a Markov chain is the expectation value at a given time of a real-valued observable on the state space.
Classically, such expectation values can be estimated efficiently by Monte Carlo sampling of the Markov chain up to the required time. In the quantum case, expectation values of quantities can be obtained by estimating the overlap of the history state with a state encoding the values of the quantity we wish to calculate.
In Appendix \ref{app:sampling}, we consider two quantum algorithms to compute such expectation values.
The first is based on a post-processing of the quantum history state of the Markov chain, and the second is based on the coherent version of the classical Monte Carlo simulation of the Markov chain. As shown in Appendix \ref{app:sampling}, one can obtain a quadratic speedup in the error of estimating an observable using quantum techniques for Monte Carlo sampling.

\paragraph*{\textbf{Post-processing: principal component analysis of data matrix}}


If the history state is effectively low rank, \textit{i.e.,} there are only a few large singular values, then the description of the time evolution of the Markov process can be compressed by expressing it in terms of the corresponding singular vectors, which the quantum principal component analysis also reveals. The history state will be effectively low rank, for example, when the Markov transition matrix has only a few eigenvalues with small negative parts, so that the dynamics is dominated over longer times by the corresponding eigenvectors. 

In the case of continuous time Markov chains, the quantum state in Eq. \ref{eq:quantumDataMatrix} can be interpreted as a $q^n \times T$ data matrix $\boldsymbol{X_{mat}}$ where each column $j$ corresponds to the probabilities of the Markov state at timestep $j$. The dominant singular values and corresponding singular vectors of this matrix can be extracted from $\ket{x}$ by performing quantum principal component analysis (qPCA) on the $\ket{x_j}$ and $\ket{j}$ registers which runs in $O(R n \log q)$ time where $R$ is the rank of $\boldsymbol{X_{mat}}$ \cite{lloyd2014quantum}. Note, that qPCA in this setting is equivalent to performing a Schmidt decomposition on the Hilbert spaces spanned by registers $\ket{x_j}$ and $\ket{j}$. The qPCA performs this decomposition via density matrix exponentiation \cite{lloyd2014quantum}. It is often the case that the effective rank $R$ (the number of large singular values) of $\boldsymbol{X_{mat}}$ is small with respect to the number of states $q^n$, and later, we show that the effective rank is in fact very small for the example models we consider. Note that because our method acts directly on a quantum state, quantum inspired algorithms for PCA do not have access to the data structure for extracting the singular vectors and singular values of the history state \cite{tang2018quantum,chia2020sampling,rudelson2007sampling}. Specifically, in Appendix \ref{app:PCA_classical}, we show that classical Monte Carlo methods cannot efficiently extract the singular vectors and the singular values of the history state (Eq. \ref{eq:quantumDataMatrix}) whenever the support of the probability distribution is exponentially large in the number of nodes of the network.

After performing qPCA via density matrix exponentiation on copies of $\ket{x}$, we have a decomposition of the data matrix into left and right singular vectors:
\begin{equation}
    \text{qPCA}: \ket{x} \rightarrow \sum_{j=0}^T \ket{l_j} \ket{r_j}\ket{\tilde \sigma_j},
    \label{eq:qSVD}
\end{equation}{}
where $\ket{l_j}$ are the left singular vectors corresponding to the Markov states,
$\ket{r_j}$ are the right singular vectors corresponding to the temporal states, and $\ket{\tilde \sigma_j}$ are estimates of the singular values.   
The singular values $\ket{\tilde \sigma_j}$ represent the weight of the left (Markov state) and right (temporal state) singular vectors in the solution.  It is conventional to take the ordering in $j$ in Eq. \ref{eq:qSVD} to be from the largest to smallest singular values.

The left singular vectors $\ket{l_j}$ can be interpreted as the most common profile of Markov states. The first left singular vector corresponds to the profile with the greatest contribution to the data matrix, often the steady state of the Markov process. The next few singular vectors typically correspond to the profile of states in the early progression of the Markov simulation before steady state is achieved (see simulations later for examples).

The right singular vectors $\ket{r_j}$ detail the progression of the corresponding left singular vectors over time. For example, the first singular vector is typically weak during the early progression and grows to a constant value as the steady state arises. The next few right singular vectors show when the corresponding left singular vectors take prominence, often highest in magnitude at early points in time (see simulations later for examples).  

Decomposing the data into singular vectors also allows one to apply efficient transformations to the singular vectors using quantum post-processing methods. For example, if one is interested in analyzing the Markov states in the frequency domain, a quantum Fourier transform can be applied to the right singular vectors. Later, in our example, we show that the dominant singular vectors correspond to slowly varying dynamics at low frequencies and the later singular vectors correspond to more rapidly varying dynamics at higher frequencies.


\paragraph*{\textbf{Post-processing: efficient quantum transformations}}
Fourier transforms and wavelet transforms are commonly used in the analysis of large datasets, especially time series \cite{lau1995climate,percival2000wavelet,cazelles2007time,grinsted2004application}. Discrete wavelet transforms, for example, can be used to identify statistical patterns in a time series. With a quantum computer, Fourier transforms and certain discrete wavelet transforms can  be performed exponentially faster than their classical counterparts \cite{hoyer1997efficient,fijany1998quantum,klappenecker2001discrete}. These transforms can be applied to the data contained in our output quantum state (Eq. \ref{eq:quantumDataMatrix}).  
For example, a Fourier transform or wavelet transform can be applied to the time register, \textit{e.g.} to observe the data in the frequency domain or to compress the data in terms of the dominant wavelets.   Let $U_{jk} = \langle k | j\rangle$ be the element of the unitary matrix $U$ that maps the states $\ket{j}$ to the transform states $\ket{k}$ (frequency states in the case of the quantum Fourier transform; wavelets in the case of the discrete wavelet transform).   Applying $U$ to the temporal register, we obtain the state
\begin{equation}
    \sum_{j=0}^T \ket{x_j} U\ket{j}  = 
    \sum_{j,k=0}^T\ket{x_j} U_{jk} \ket{k} =
    \sum_{k=0}^T \ket{y_k}\ket{k},
    \label{eq:quantumtransform}
\end{equation}
where $\ket{y_k} = \sum_{j=0}^T U_{jk} \ket{x_j}$, is the state of the system correlated with the $k$th frequency or wavelet state in the temporal register.   Sampling from the temporal register then yields the dominant frequencies/wavelets, and the spatial register yields the state of the system correlated with those frequencies/wavelets. For example, as noted above in the discussion of the Schr\"{o}dinger equation, in Eq. \ref{eq:generalDiff}, if $\vec c$ is $0$ and the matrix ${\cal M}$ is anti-Hermitian, performing a quantum Fourier transform on the temporal register yields the purely imaginary eigenvalues of ${\cal M}$ and the output contains the corresponding eigenvectors \cite{shao2019computing}.   More generally, when the eigenvalues of ${\cal M}$ have both real and imaginary components, performing the quantum Fourier transform
reveals the power spectrum of the solution: a complex eigenvalue $a + ib$ manifests itself as a Lorentzian peaked at the natural frequency $\sqrt{a^2 + b^2}$. 

By contrast, classical Monte Carlo sampling does not obviously extract the proper information for performing Fourier or wavelet transforms on the quantum state (see Appendix \ref{app:PCA_classical}).

\paragraph*{\textbf{Post-processing: quantum machine learning}}
In the past few years, many quantum algorithms for machine learning have been proposed that can be performed exponentially faster than classical counterparts when data inputs are quantum states \cite{biamonte2017quantum,rebentrost2014quantum,li2015experimental,schuld2016prediction,wang2017quantum,farhi2018classification,grilo2019learning,romero2017quantum}. 
When data in the form of Eq. \ref{eq:quantumDataMatrix} is used as input, applications of machine learning algorithms are numerous. Here we list some of these applications, grouped by the type of model used. First, quantum models have been proposed for compression of data or efficient readout. These models include quantum auto-encoders \cite{romero2017quantum,khoshaman2018quantum} and qPCA as discussed before \cite{lloyd2014quantum}. Second, a wide range of algorithms implementing kernel methods can be used to classify data, identify key features in the data, or measure similarity between different datasets \cite{rebentrost2014quantum,schuld2019quantum,havlivcek2019supervised,lloyd2020quantum}.   Indeed, if we trace out the temporal register in Eq. \ref{eq:quantumDataMatrix}, the state register is described by the (unnormalized) density matrix $\sum_j \ket{x_j}\bra{x_j}$, which is the covariance matrix for the synthetically generated data which can be analyzed directly.
Third, output states can be input into parameterized quantum circuits or quantum neural networks to identify key features or perform machine learning tasks \cite{cong2019quantum,farhi2018classification,schuld2020circuit}.

\subsection{Example simulation for epidemic processes}
\label{sec:main_example}
The analysis of epidemic spreading -- viral or social -- is often modeled as a dynamical process on a complex network \cite{pastor2015epidemic}. Theoretical approaches to epidemic processes typically assume transitions (\textit{e.g.} rates of infection) occur as Poisson processes which correspond to models of continuous time Markov chains \cite{pastor2015epidemic,van2012epidemics}. Classically, numerical simulation of continuous time Markov processes is intractable for large networks as the dimension of the Markov state grows exponentially with the number of nodes in the network or graph. 

To demonstrate the applicability of our quantum algorithm, we simulate continuous time Markov chain models on simple seven node networks.  We choose a relatively small network so that we can still visually represent the full solution to the continuous time Markov chain. Here, we present models for analyzing viral epidemics and perform similar analysis for social epidemics in Appendix \ref{app:simulations}. For ease of graphical presentation, we implement susceptible-infected-susceptible (SIS) epidemiological models which have only two states per node: susceptible and infected. This is in contrast to the more realistic susceptible-infected-recovered (SIR) models which also fits within the framework of our algorithms, but can be hard to visualize and plot since they have many more states ($3^n$ as opposed to $2^n$). 

\begin{figure*}[ht]
\centering
\includegraphics[width=3.5in]{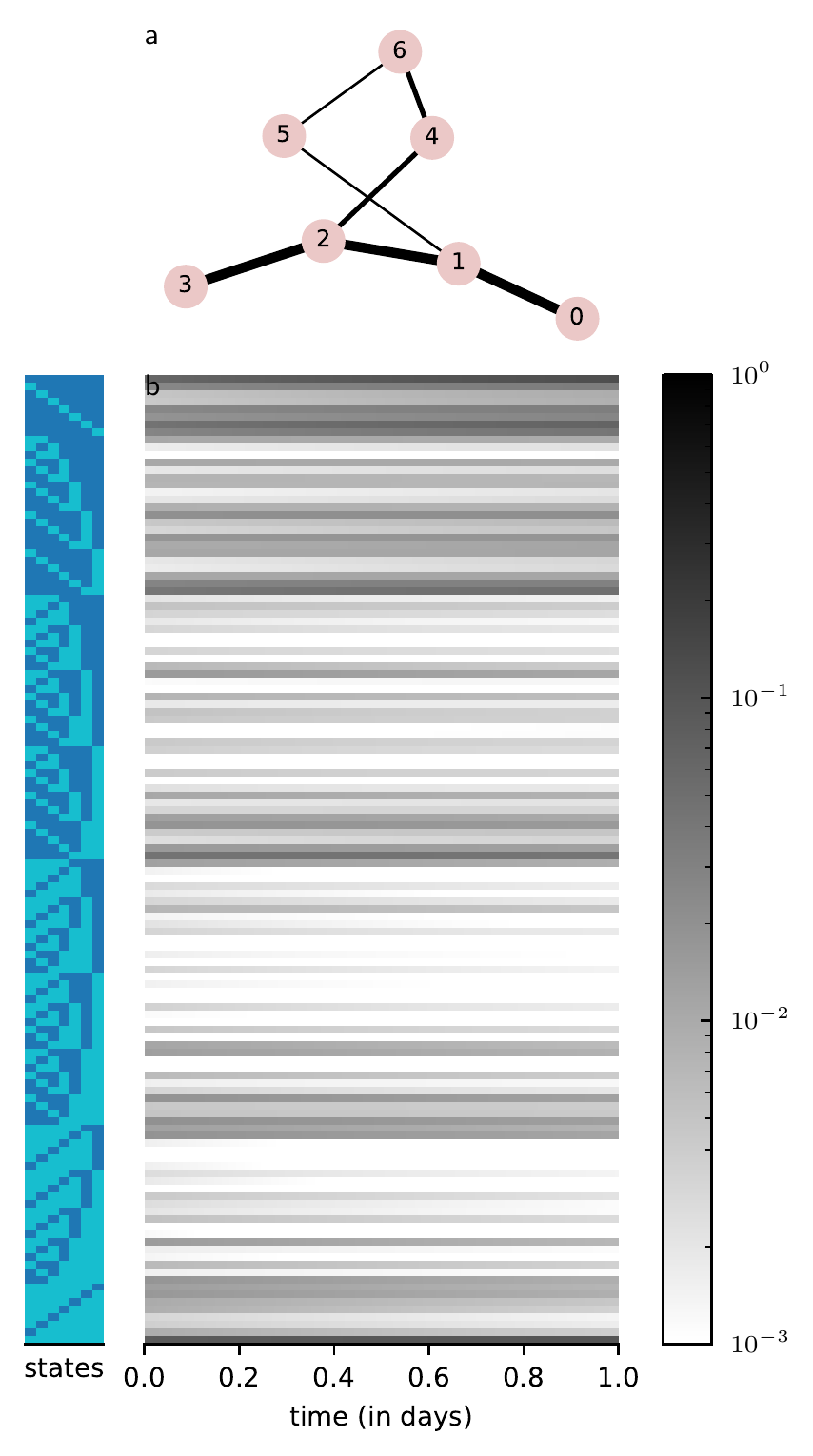}
\caption{ \textbf{(a)} The 7 node network used for simulation of the continuous time Markov chain. Line widths correspond to the rate of infection from infected neighbor nodes ($r_{SI} = \{ 0.4, 0.8, 1.6 \}$ from thinnest to thickest lines). Network has the feature that some nodes (\textit{e.g.,} node 2 and 3) are strongly connected to neighboring nodes and others  (\textit{e.g.,} node 5) are not. \textbf{(b)} Probabilities of Markov states over the course of three days shown as a colorbar chart (logarithmically scaled). Post-processing is performed on the data between days one and two contained in the orange box. States are enumerated as rows on the left hand side, each denominated by a 7 node colorbar numbered node 0 on the left to node 6 on the right. Dark/light color indicates a node in that state is infected/susceptible respectively.}
\label{fig:main-network}
\end{figure*}

\paragraph*{\textbf{Epidemic simulations of viral contagion}}
We present analysis of a Markov simulation for a susceptible-infected-susceptible (SIS) model on a single network shown in Fig. \ref{fig:main-network}a. Our simple model features many common properties of continuous time Markov chain simulations. Notably, it is common that Markov transition matrices have only a small number of dominant eigenvalues (\textit{i.e.,} those whose values are close to zero) thus rendering them suitable for analysis similar to that performed here for small networks. Of course, network models with more nodes will likely have emergent phenomena that will not appear in this small network -- phenomena that one may hope to analyze using quantum computers \cite{pastor2015epidemic,van2012epidemics}.

For an SIS model, the full state at time $t$ is described by a vector $\vec x(t)$ of length $2^n$ ($n=7$ for our example). All transitions are modeled as Poisson processes with transition matrix $\boldsymbol{Q}$. 
\begin{equation}
    \frac{d \vec x(t)}{dt} = \boldsymbol{Q} \vec x(t)
\end{equation}{}

We begin in an initial state where any node has a 35\% probability of being infected and conduct analysis over the intermediate phase of the epidemic, between days one and two. To numerically estimate the state at discretized times $\vec x_t$, we employ a Forward Euler method that begins at the state of the epidemic on day one and ends at day two over $T=1027$ timesteps (step size $h \approx 0.001$ days):
\begin{equation}
    \vec x_{t+1} = \vec x_t + h \boldsymbol{Q} \vec x_t
\end{equation}
From day one to day two, the epidemic has spread sufficiently that multiple individuals are likely to be infected, and the probability distribution has spread throughout the space.  As noted above, this is the regime where the quantum algorithm provides an exponential advantage over existing classical Monte Carlo techniques.

Fully simulating the above for $T$ steps constructs a data matrix $\boldsymbol{X_{mat}}$ where column $i$ is equal to $x_{i-1}$ (we assume initial state $x_0$ is included in this matrix as well). 
\begin{equation}
\boldsymbol{X_{mat}} = 
\begin{bmatrix}
    \vert & \vert & & \vert  \\
    \vec x_0   & \vec x_1 & \dots & \vec x_T \\
    \vert & \vert & & \vert
\end{bmatrix}
\end{equation}
where we note that this data matrix can be interpreted as a classical version of the quantum output state shown in Eq. \ref{eq:quantumDataMatrix}.

In Fig. \ref{fig:main-network}b, we plot the probabilities of states of the Markov process over time providing a visualization of the data matrix $\boldsymbol{X_{mat}}$. Subsequent post-processing of this data matrix is performed on the segment of data between days one and two. We note that though Fig. \ref{fig:main-network}b contains a complete description of the Markov history state, extracting trends and analyzing this figure can be challenging. For larger networks, visualization of this data matrix cannot be efficiently performed and we now turn our attention to efficient quantum algorithms for extracting salient features from this data matrix.

\begin{figure*}[t]
    \centering
    \includegraphics{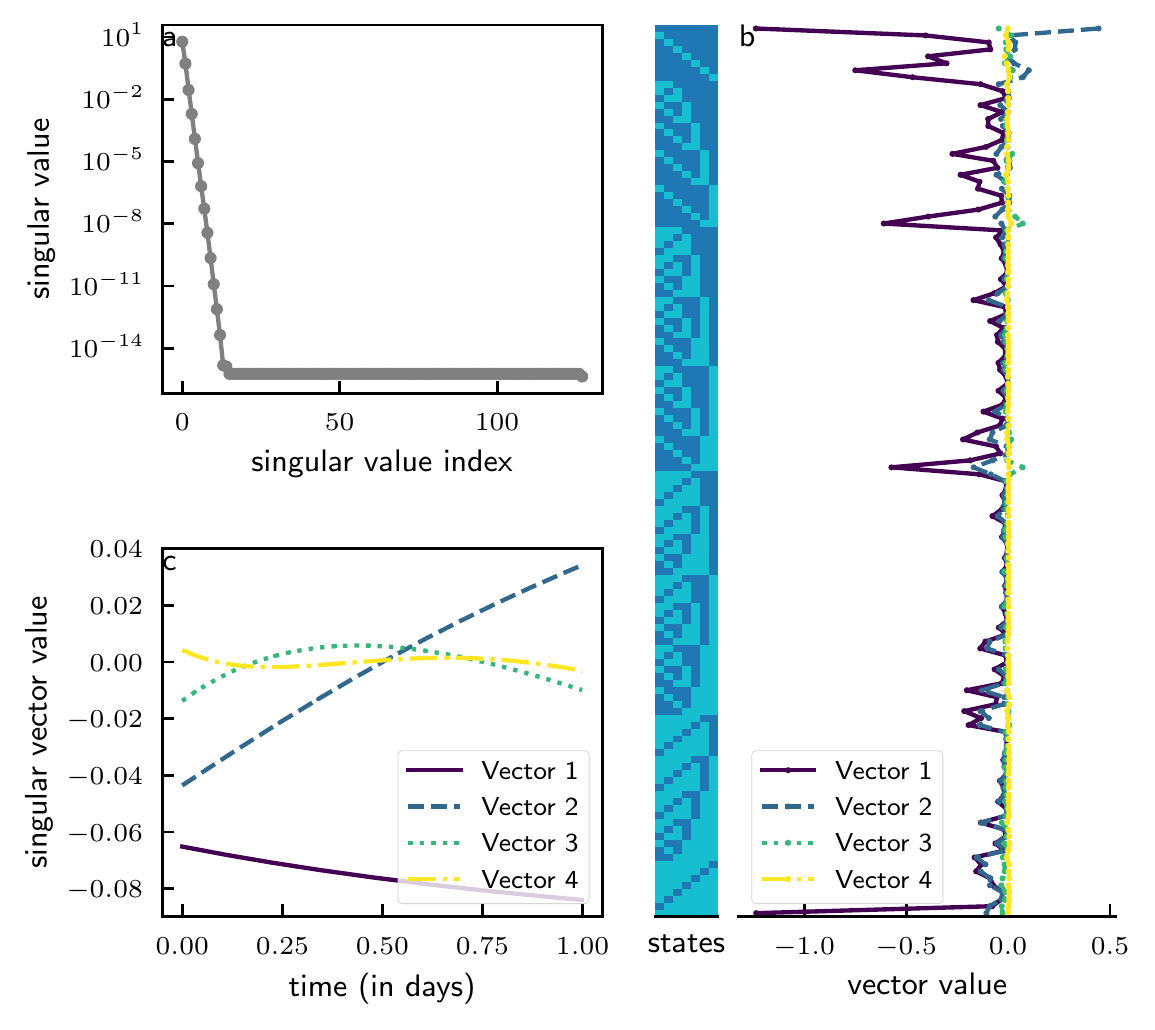}
    \caption{\textbf{(a)} Singular values of data matrix $\boldsymbol{X_{mat}}$ decay exponentially fast. \textbf{(b)} First four left singular vectors scaled by the square root of their corresponding singular values show that much of the disease progression can be understood by just observing the first few vectors. States are enumerated as rows on the left hand side, each denominated by a 7 node colorbar numbered node 0 on the left to node 6 on the right. Dark/light colors indicate a node in that state is infected/susceptible respectively. \textbf{(c)} Values of the right singular vectors scaled by the square root of their corresponding singular values show the progression of the epidemic over time. The first singular vector depicts the steady state and the next few singular vectors detail the intermediate course of the Markov process. }
    \label{fig:main-singular}
\end{figure*}

\paragraph*{\textbf{Simulations: quantum principal component analysis (qPCA)}}
The output of our continuous time Markov chain algorithm is stored in a data matrix, visualized in Fig. \ref{fig:main-network}b. In the quantum setting, this data matrix is a quantum state which we can subsequently post-process using various efficient quantum algorithms. One available post-processing algorithm is quantum Principal Component Analysis (qPCA) as in Eq. \ref{eq:qSVD}, which can compress the data into its singular vectors \cite{lloyd2014quantum}. If the matrix is low rank, as in our example with exponentially decaying singular values (see Fig. \ref{fig:main-singular}a), qPCA can be performed in time logarithmic in the dimension of the full Markov state \cite{lloyd2014quantum}. The principal components can be subsequently transformed or even measured.

The most dominant left (Fig. \ref{fig:main-singular}b) and right (Fig. \ref{fig:main-singular}c) singular vectors can be interpreted as the most common profile of states (left vectors) and their corresponding trajectories in time (right vectors). In our example, the first singular vector corresponds to the steady state of our epidemic. Note that it takes prominence almost completely throughout the course of the simulation (see right singular vector in Fig. \ref{fig:main-singular}c). The second singular vector plots important changes in the epidemic as more nodes become infected. The corresponding right singular vector plots the steady, almost linear, transition over time as this singular vector takes prominence. Similarly, the third and fourth singular vectors plot trends in the progression of the epidemic, especially in early phases where nodes become infected.

\begin{figure}[ht]
    \centering
    \includegraphics{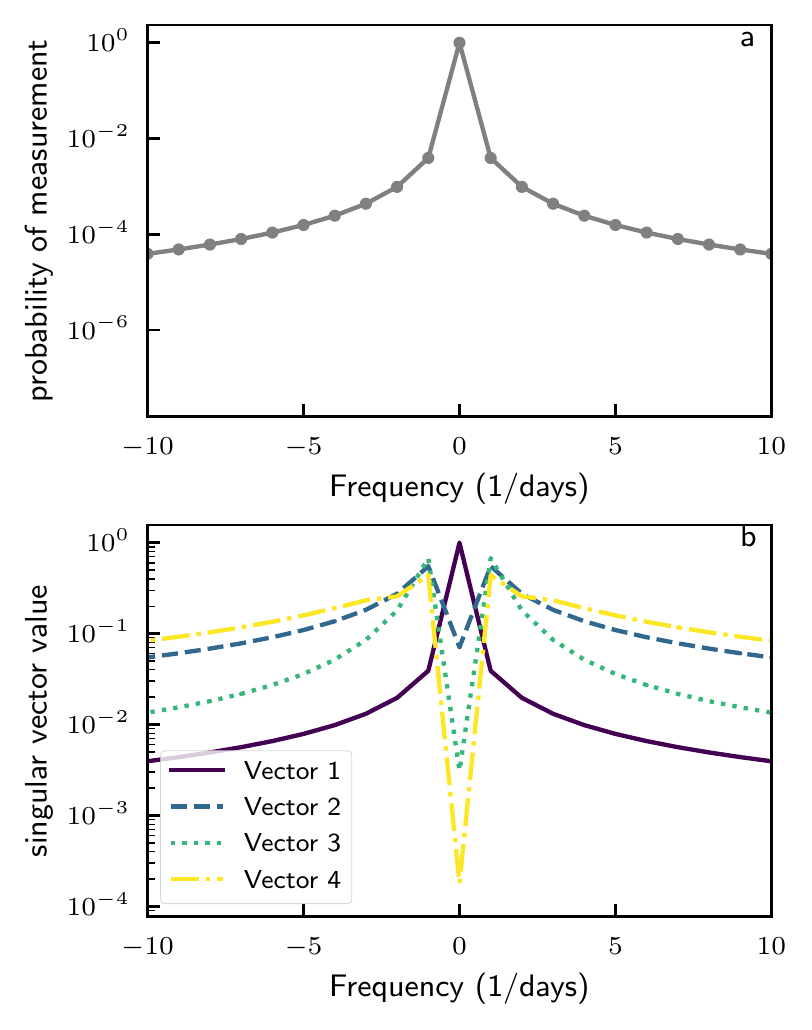}
    \caption{\textbf{(a)} Measurements made in the Fourier domain (plotted only for low frequency components) have high probability in the zero frequency (steady state) and low frequency components. \textbf{(b)} The first singular vector, plotted in the Fourier domain, is dominated by the zero frequency component corresponding to the steady state. Later singular vectors take prominence in low frequency components which indicate the rate of change during the intermediate progression of the Markov chain. }
    \label{fig:main_fft}
\end{figure}

\paragraph*{\textbf{Simulations: Fourier and wavelet analysis}}
For small networks such as the one studied here, the data in the Fourier domain is dominated by the steady state contributions as shown in Fig. \ref{fig:main_fft}a. With quantum algorithms, one also has the option of transforming the individual singular vectors into their frequency components as shown in Fig. \ref{fig:main_fft}b. The second to fourth singular vectors all have strong contributions from low frequencies, whose values provide an indication of the rate of change in the progression of the Markov chain. Given the small network size, this dominance of low frequency components is perhaps not altogether surprising. Larger networks encounter phenomena not observed in small networks, and may potentially reveal interesting features in the Fourier domain \cite{pastor2015epidemic,van2012epidemics} if they are analyzed with a quantum computer.

\begin{figure*}[ht]
    \centering
    \includegraphics{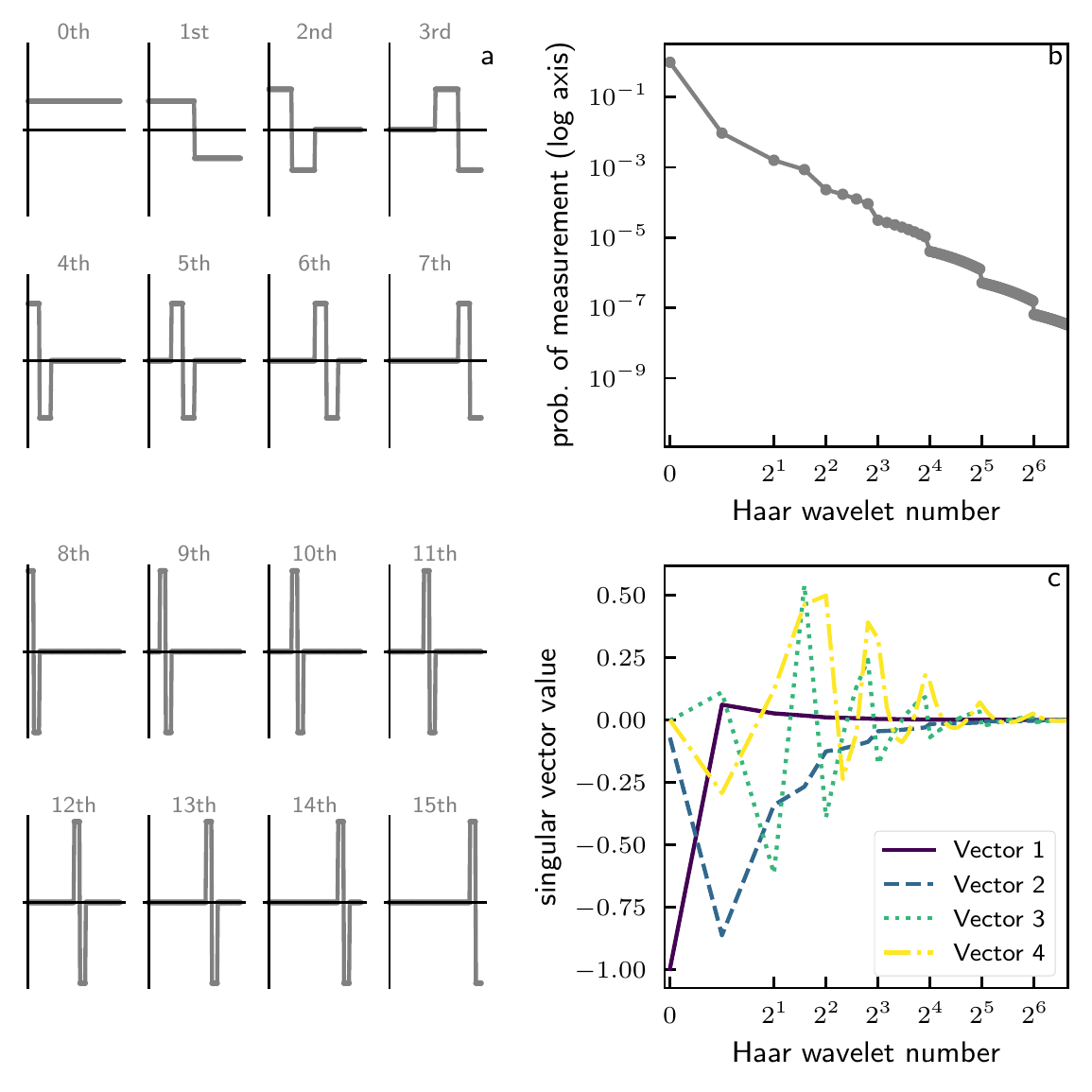}
    \caption{\textbf{(a)} Each Haar wavelet has support over a characteristic timescale indicated by the wavelet number. Haar wavelet numbers between powers of $2$ take the same form and are offset from each other in the time dimension. As a visual aid, we show here the first 16 discrete Haar wavelets. \textbf{(b)} Quantum state has high probability in the zeroth Haar wavelet (steady state). \textbf{(c)} Wavelet transform of singular vectors shows that the first singular vector is strongest in the zeroth Haar vector (steady state). Later singular vectors take prominence in Haar vectors with support in the early stages of epidemic. }
    \label{fig:main_haar}
\end{figure*}

Beyond the standard Fourier transform, quantum computers offer the advantage of efficient post-processing via wavelet transforms \cite{hoyer1997efficient}. Continuing the example shown in the main text, here we transform the time dimension of our data matrix using a Haar wavelet transform, which can be performed efficiently on a quantum computer \cite{hoyer1997efficient}. When viewed in the Haar wavelet basis (form of wavelets shown in Fig. \ref{fig:main_haar}a), one can analyze the characteristic timescales over which differences in the Markov state probabilities become apparent. Perhaps unsurprisingly, as shown in Fig. \ref{fig:main_haar}b, the zeroth Haar vector is most prominent as this corresponds to the steady state of the Markov chain. More interesting results are observed in analysis of the singular vectors, which can also be transformed into the Haar domain as shown in Fig. \ref{fig:main_haar}c. Here, clear differences can be observed in the Haar basis of the steady state singular vector (first singular vector) and later vectors. The first singular vector is dominated by the zeroth Haar wavelet (constant wavelet) since that singular vector corresponds to the steady state. The next few singular vectors corresponding to changes in the intermediate progression of the epidemic are dominated by Haar vectors with support over various phases. For example, the third and fourth singular vectors take large values over Haar vectors with support in the early phases of the simulation (\textit{e.g., } fourth and eighth Haar vectors).

\subsection{Potential for realization on near term quantum devices}
The algorithms proposed here are potentially suitable for near term quantum devices \cite{preskill2018quantum} with around 100 qubits. We note that the presence of noise in near term quantum devices, which is not analyzed here, may present a challenge to the successful implementation of these algorithms. Nevertheless, if challenges with noise are addressed via error correction or other means, our algorithms can be used to analyze the output of high dimensional differential equations outside the reach of available classical algorithms. For example, Markov states of dimensionality $2^{50}$ can be simulated for a million timesteps (approximately $2^{20}$) on a quantum computer with 100 qubits using algorithms from \cite{berry2017quantum}. Given the output state, quantum principal component analysis and wavelet transforms can subsequently be performed to analyze the history state generated by the near term quantum device \cite{lloyd2014quantum,bellante2021quantum,hoyer1997efficient}.

\section{Discussion}
Common quantum algorithms for solving linear differential equations output quantum states corresponding to solutions of physical models in high dimensional vector spaces. These output states store the complete history of the solution to a differential equation, allowing one to perform efficient quantum post-processing on these solution states.  Our approach avoids a commonly cited drawback of many quantum machine learning and data processing algorithms -- that classical inputs must be mapped into quantum states or stored in qRAM.  
We focus here on the case of Markov models and propose efficient quantum algorithms for evaluating continuous time Markovian and non-Markovian models. Our algorithms allow for efficient simulation of Markov models on the complete state of a Markov chain. Outputs of our models are quantum states which can be efficiently generated and then passed as inputs for other efficient quantum post-processing algorithms  (\textit{e.g.,} quantum signal analysis and machine learning).  The quantum post-processing reveals features of the data such as the singular value decomposition, the power spectrum, and wavelet decompositions, which cannot be reconstructed efficiently using classical sampling algorithms.  When applied to complex networks, the quantum algorithms may be used to reveal such fundamental features of the dynamics of epidemics potentially exponentially faster than classical algorithms.

\paragraph*{\textbf{Data Availability}}
All data generated or analysed during this study are included in this article (and supplementary information files). 

\paragraph*{\textbf{Code Availability}}
To recreate figures and results, please access our code repository at:\\ \href{https://github.com/bkiani/Quantum-DiffEQ-Advantage}{https://github.com/bkiani/Quantum-DiffEQ-Advantage}

\section*{Acknowledgements}
This work was supported by NSF, ARO, DOE, AFOSR, and IARPA. MM is supported by the NSF Grant No.~CCF-1954960.  The authors thank Peter Love and Clemens Wittenbecher for helpful discussions.

\section*{Author Contributions}
All authors designed the research and analyzed results. BK and GP performed the research including the mathematical proofs and computations. BK, GP, and SL wrote the paper.


\bibliographystyle{apsrev4-2}
\bibliography{main}

\appendix
\begin{appendices}
\onecolumngrid

\section{Proofs of runtimes and error bounds} \label{app:runtime_proofs}
\subsection{Generation of the quantum history state}
Let $T = \left\lceil t_{\max}\left\|\mathcal{M}\right\|\right\rceil$, let $\delta>0$ and let $k\in\mathbb{N}$ with $k\ge5$ and $(k+1)!\ge2T$.
The quantum algorithm of Ref. \cite{berry2017quantum} discretizes the differential equation in $T$ time steps of size $h = t_{\max}/T$, truncates the Taylor expansion of the matrix exponential at the $k$-th order and produces an approximate normalized version $|\phi\rangle$ of the quantum state
\begin{equation}
|y\rangle = \sum_{i=0}^{T-1}\sum_{j=0}^k|i,j\rangle|y_{i,j}\rangle + |T,0\rangle|y_{T,0}\rangle\,.
\end{equation}
For sufficiently large $k$, the vectors $|y_{i,0}\rangle$ are close to the solution of the linear differential equation at the $i$-th time step \cite[Theorem 6]{berry2017quantum}:
\begin{equation}\label{eq:errori}
\left\|\vec{x}(ih) - |y_{i,0}\rangle\right\|\le 2.8\,\kappa\,i\,\frac{\left\|\vec{x}(0)\right\|+t_{\max}\left\|\vec{c}\right\|}{\left(k+1\right)!}\,,
\end{equation}
and the vectors $|y_{i,j}\rangle$ for $j\ge1$ are some garbage vectors associated to the terms of the Taylor expansion of the matrix exponential, of which we do not need the exact form.
Their norms are upper bounded by \cite[eqs. (99), (100)]{berry2017quantum}
\begin{equation}\label{eq:normij}
\left\||y_{i,j}\rangle\right\| \le \frac{\left\||y_{i+1,0}\rangle\right\| + \left\||y_{i,0}\rangle\right\|}{\left(3-\mathrm{e}\right)j!}\,.
\end{equation}
Let
\begin{equation}
|\bar{y}\rangle = \frac{|y\rangle}{\left\||y\rangle\right\|}
\end{equation}
be the normalized version of $|y\rangle$.
The quantum state $|\phi\rangle$ satisfies
\begin{equation}\label{eq:delta}
    \left\||\phi\rangle - |\bar{y}\rangle\right\| \le \delta
\end{equation}
and the algorithm requires
\begin{equation}\label{eq:complC}
O\left(\kappa\,k^2\,t_{\max}\left\|\mathcal{M}\right\|s\,\mathrm{poly}\log\frac{\kappa\,k\,t_{\max}\left\|\mathcal{M}\right\|s\,N}{\delta}\right)
\end{equation}
elementary quantum gates \cite[eq. (128)]{berry2017quantum}.

An approximation of the quantum history state $|x\rangle$ can be obtained by projecting the second register of the quantum state $|\phi\rangle$ on the $0$ value.
Let $\langle0|\phi\rangle$ be the unnormalized projection.
In the following, we show that its success probability is $O(1)$, and that choosing
\begin{align}
\delta &= O(\epsilon)\,,\nonumber\\
k &= O\left(\log\left(\left(1+\frac{t_{\max}\left\|\vec{c}\right\|}{\left\|\vec{x}(0)\right\|}\right)\frac{\kappa\,t_{\max}\left\|\mathcal{M}\right\|}{\epsilon}\right)\right)
\end{align}
we can achieve
\begin{equation}
    \left\|\frac{\langle0|\phi\rangle}{\left\|\langle0|\phi\rangle\right\|} - |\bar{x}\rangle\right\|\le\epsilon\,.
\end{equation}
From Eq. \ref{eq:complC}, this modification of the algorithm of Ref. \cite{berry2017quantum} requires
\begin{align}
&O\left(\kappa\,t_{\max}\left\|\mathcal{M}\right\|s\right.\nonumber\\
&\qquad \left.\mathrm{poly}\log\left(\left(1+\frac{t_{\max}\left\|\vec{c}\right\|}{\left\|\vec{x}(0)\right\|}\right)\frac{\kappa\,t_{\max}\left\|\mathcal{M}\right\|s\,N}{\epsilon}\right)\right)
\end{align}
elementary quantum gates.

\subsubsection*{Success probability}
We assume that
\begin{equation}\label{eq:hdelta}
    \delta \le \frac{1}{2\sqrt{66}}\,.
\end{equation}
We have from Eq. \ref{eq:normij}
\begin{align}
\sum_{j=1}^k\left\||y_{i,j}\rangle\right\|^2 &\le \frac{\left(\left\||y_{i+1,0}\rangle\right\| + \left\||y_{i,0}\rangle\right\|\right)^2}{\left(3-\mathrm{e}\right)^2}\sum_{j=1}^\infty \frac{1}{{j!}^2}\nonumber\\
&\le 1.28*2\,\frac{\left\||y_{i+1,0}\rangle\right\|^2 + \left\||y_{i,0}\rangle\right\|^2}{\left(3-\mathrm{e}\right)^2}\,,
\end{align}
and
\begin{align}
\sum_{i=0}^{T-1}\sum_{j=1}^k\left\||y_{i,j}\rangle\right\|^2 &\le \frac{1.28*4}{\left(3-\mathrm{e}\right)^2}\sum_{i=0}^{T}\left\||y_{i,0}\rangle\right\|^2\nonumber\\
&\le 65\sum_{i=0}^{T}\left\||y_{i,0}\rangle\right\|^2\,.
\end{align}
Let $\langle0|y\rangle$ be the projection of $|y\rangle$ onto the $0$ value of the second register.
The success probability of such projection is
\begin{align}\label{eq:py}
\left\|\langle0|\bar{y}\rangle\right\|^2 &= \frac{\sum_{i=0}^{T}\left\||y_{i,0}\rangle\right\|^2}{\sum_{i=0}^{T}\left\||y_{i,0}\rangle\right\|^2 + \sum_{i=0}^{T-1}\sum_{j=1}^k\left\||y_{i,j}\rangle\right\|^2}\nonumber\\
&\ge \frac{1}{66}\,.
\end{align}
Let $\langle0|\phi\rangle$ be the projection of $|\phi\rangle$ onto the $0$ value of the second register.
We have from Eq. \ref{eq:delta}, \ref{eq:py} and \ref{eq:hdelta}
\begin{align}
    \left\|\langle0|\phi\rangle\right\| &\ge \left\|\langle0|\bar{y}\rangle\right\| - \left\|\langle0|\bar{y}\rangle - \langle0|\phi\rangle\right\|\nonumber\\
    &\ge \left\|\langle0|\bar{y}\rangle\right\| - \delta \ge \frac{1}{2\sqrt{66}}\,,
\end{align}
therefore the success probability of the projection satisfies
\begin{equation}
    p = \left\|\langle0|\phi\rangle\right\|^2 \ge \frac{1}{264}\,.
\end{equation}

\subsubsection*{Error analysis}
From Eq. \ref{eq:errori}, the distance between the quantum history state $|x\rangle$ and the projection $\langle0|y\rangle$ satisfies
\begin{align}
&\left\||x\rangle - \langle0|y\rangle\right\|^2 = \sum_{i=0}^T\left\|\vec{x}(ih) - |y_{i,0}\rangle\right\|^2\nonumber\\
&\le 2.8^2\,\kappa^2\frac{\left(\left\|\vec{x}(0)\right\|+t_{\max}\left\|\vec{c}\right\|\right)^2}{{\left(k+1\right)!}^2} \frac{T\left(T+1\right)\left(2T+1\right)}{6}\,,
\end{align}
then
\begin{equation}
\left\||x\rangle - \langle0|y\rangle\right\| \le  \kappa\frac{\left\|\vec{x}(0)\right\|+t_{\max}\left\|\vec{c}\right\|}{2^{k+4}} \sqrt{3T}\left(T+1\right)\,.
\end{equation}
We have from Lemma \ref{lem:norm}
\begin{align}
\left\||\bar{x}\rangle - \frac{\langle0|y\rangle}{\left\|\langle0|y\rangle\right\|}\right\| &\le \frac{\left\|\vec{x}(0)\right\|+t_{\max}\left\|\vec{c}\right\|}{\left\||x\rangle\right\|} \frac{\kappa\sqrt{3T}\left(T+1\right)}{2^{k+3}}\nonumber\\
&\le \left(1+\frac{t_{\max}\left\|\vec{c}\right\|}{\left\|\vec{x}(0)\right\|}\right)\frac{\kappa\sqrt{3T}\left(T+1\right)}{2^{k+3}}\,.
\end{align}
We choose
\begin{align}
    k &= \left\lceil\log_2\left(\left(1+\frac{t_{\max}\left\|\vec{c}\right\|}{\left\|\vec{x}(0)\right\|}\right)\frac{\kappa\sqrt{3T}\left(T+1\right)}{4\epsilon}\right)\right\rceil\nonumber\\
    &= O\left(\log\left(\left(1+\frac{t_{\max}\left\|\vec{c}\right\|}{\left\|\vec{x}(0)\right\|}\right)\frac{\kappa\,t_{\max}\left\|\mathcal{M}\right\|}{\epsilon}\right)\right)\,,
\end{align}
such that
\begin{equation}
\left\||\bar{x}\rangle - \frac{\langle0|y\rangle}{\left\|\langle0|y\rangle\right\|}\right\| \le \frac{\epsilon}{2}\,.
\end{equation}
We have from Eq. \ref{eq:delta}, \ref{eq:py} and Lemma \ref{lem:norm} again
\begin{equation}
\left\|\frac{\langle0|\phi\rangle}{\left\|\langle0|\phi\rangle\right\|} - \frac{\langle0|\bar{y}\rangle}{\left\|\langle0|\bar{y}\rangle\right\|}\right\| \le \frac{2\left\|\langle0|\phi\rangle - \langle0|\bar{y}\rangle\right\|}{\left\|\langle0|\bar{y}\rangle\right\|} \le  2\sqrt{66}\,\delta\,,
\end{equation}
such that choosing
\begin{equation}
    \delta = \frac{\epsilon}{4\sqrt{66}}
\end{equation}
we have
\begin{equation}
\left\|\frac{\langle0|\phi\rangle}{\left\|\langle0|\phi\rangle\right\|} - \frac{\langle0|\bar{y}\rangle}{\left\|\langle0|\bar{y}\rangle\right\|}\right\| \le \frac{\epsilon}{2}
\end{equation}
and
\begin{align}
    \left\|\frac{\langle0|\phi\rangle}{\left\|\langle0|\phi\rangle\right\|} - |\bar{x}\rangle\right\| &\le \left\|\frac{\langle0|\phi\rangle}{\left\|\langle0|\phi\rangle\right\|} - \frac{\langle0|\bar{y}\rangle}{\left\|\langle0|\bar{y}\rangle\right\|}\right\| \nonumber\\
    &\phantom{\le} + \left\|\frac{\langle0|y\rangle}{\left\|\langle0|y\rangle\right\|} - |\bar{x}\rangle\right\| \le \epsilon
\end{align}
as required.

\begin{lem}\label{lem:norm}
Let $v,\,w$ be vectors in a normed vector space.
Then,
\begin{equation}
\left\|\frac{v}{\|v\|} - \frac{w}{\|w\|}\right\| \le 2\,\frac{\|v-w\|}{\|v\|}\,.
\end{equation}
\begin{proof}
We have
\begin{equation}
\left\|\frac{v}{\|v\|} - \frac{w}{\|w\|}\right\| \le \frac{\|v-w\| + \left|\|w\|-\|v\|\right|}{\|v\|} \le 2\,\frac{\|v-w\|}{\|v\|}\,.
\end{equation}
\end{proof}
\end{lem}

\section{Input state preparation}
\label{app:input_preparation}
Input states to differential equations encode boundary conditions and initial states. These input states can be prepared via two different methods discussed here.

\paragraph{Efficiently constructed quantum initial states}
The most optimal setting which commonly occurs in differential equation analysis is when input vectors are sparse or efficient to construct on a quantum computer. For a standard linear differential equation in the form $\frac{d \vec x(t)}{dt} = {\cal M} \vec x(t) + \vec c$, the boundary conditions take the form below (copied from the main text):
\begin{equation*}
  \vec b =  
  \begin{pmatrix} 
              \vec x_0 \\ 
              \vec c \\ 
              \vdots \\ 
              \vec c
  \end{pmatrix}.
\end{equation*}

Note, that in quantum algorithms, copies of the vector above are encoded into the quantum state at appropriate locations. It is often the case that $\vec b$ is sparse or easily computed by local operations. This is especially true in the case of Markov chains algorithms where $c=0$ and the initial state is either sparsely supported or locally independent. For example, applying Hadamard gates to all qubits in the initial state constitutes one easy method to construct the initial state with uniform support over all states. Similarly, applying single qubit operations to each node provides an easy method to form an initial state where one has knowledge of individual nodes independent of other nodes.

\paragraph{Input states via qRAM}
If it is not possible to efficiently construct the initial state via sparse matrix or local operations as discussed above, another option one can use is inputting states via a quantum random access memory (qRAM) data structure \cite{kerenidis2016quantum,kerenidis2020quantum}. qRAM data structures store data $(i, x_i) \in [n]$ in a form that allows for quantum queries in superposition $\ket{i,0} \to 
\ket{i,x_i}$ in time $O(\text{poly} \log (n))$. Constructing such a data structure would, in general, require $O(n)$ quantum operations \cite{kerenidis2020quantum}. However, in settings where subsequent computations require longer runtimes, say $O(n^2)$ or $O(n^3)$ time, then such a qRAM data structure can be efficiently employed in our algorithms.

\section{Sampling and calculating observables of Markov states}
\label{app:sampling}

To determine the $T q^n$ individual probabilities of a Markov state $x_j(k)$, for $j=0$ to $T$, $k=1$ to $ N = q^n$ stored in a quantum state, one would have to make an exponentially large number of measurements. However, one can use quantum measurements to reveal a wide
variety of desired properties of the quantum system at time
$j$.  To extract the expectation
value of some desired quantity $Q$ (total number infected, variance of the 
infection rate across the graph, existence of hot spots, etc.)
we need to make a quantum measurement to estimate the expectation
value 
\begin{equation}
    \langle Q \rangle = \sum_{k=1}^N x_j(k) Q(k)
\end{equation}
where $Q(k)$ is the value of $Q$ on the $k$'th state of the vertices
of the graph.

First, we have to make sure that we obtain the state $|\vec x_j\rangle$
with high probability.   In the formulation given above, this state
only occurs in the overall superposition $|x\rangle$ with amplitude
$ O(1/\sqrt{T})$.      The standard way to amplify the probability
of finding the answer at the desired time $j$ \cite{berry2014high} is to pad out 
the matrix $A$ following the $j$'th row with $O(T)$ rows of the form 
\begin{equation}
    ( 0 \ldots 0 ~  - I ~ I ~0 \ldots 0)
\end{equation}
where the $I$ term in each $ - I ~ I ~ 0$ sequence lies on the diagonal.
These rows induce a trivial dynamics in which all the states in the 
solution following the $j$'th state also contain the state $|\vec x_j\rangle$.
This technique allows us to obtain the state $|\vec x_j\rangle$
with probability $O(1)$.  

To obtain $\langle Q \rangle = \sum_{k=1}^N x_j(k) Q(k)$,
we use standard techniques of quantum state preparation \cite{grover2002creating}.
We assume that we are given access in quantum superposition to
the individual values of the variable $Q(k)$ together with
its partial sums over ranges of $k$;   The techniques of \cite{grover2002creating}
then allow us to construct the (unnormalized) state 
\begin{equation}
    | Q\rangle = \sum_{k=1}^N Q(k) |k\rangle
\end{equation}
in time $O(\log N) = O(n\log q)$.
Define $Z_Q = \sum_{k=1}^N Q(k)^2$.  Even if $Z_Q$ is not known
beforehand, its value is revealed during the state preparation
process \cite{grover2002creating}.   
The normalized version of $|Q\rangle$ is then
\begin{equation}
    |\tilde Q \rangle = Z^{-1/2}_Q |Q\rangle
\end{equation}

We can now use a swap test between $|\tilde Q\rangle$ and $|\tilde x_j\rangle$
to measure the overlap $\langle\tilde Q| \tilde x_j\rangle$.   This overlap in 
turn allows us to measure 
\begin{equation}
    \langle Q \rangle = \sum_{k=1}^N x_j(k) Q(k) = \langle Q| x_j\rangle
= Z_j^{1/2} Z_Q^{1/2} \langle \tilde Q| \tilde x_j\rangle
\end{equation}
 
In conclusion, even though we don't have access to the individual
probabilities for states, the quantum algorithm allows us
to measure expectation values for a wide variety of observables
efficiently. This method allows us to use the quantum algorithm to
extract expectation values of the desired quantities.

\paragraph*{Comparison to classical complexity}
Extracting expectation values could also be done classically using
classical Monte Carlo to sample from the probabilistic dynamics.
Because of the local form of the probabilistic updating rule, the number
of computational steps required to draw one sample of the Markov chain at time $t$ scales as
\begin{equation}\label{eq:compls}
    O(n (t/h) \log q)\,.
\end{equation}
The average of $Q$ over $O\left(\left.\log\frac{1}{\delta}\right/\epsilon^2\right)$ samples is $\epsilon$-close to $\langle Q\rangle$ with probability at least $1-\delta$, and its computation has complexity
\begin{equation}\label{eq:classicalscaling}
O\left(\frac{n (t/h)\log\frac{1}{\delta}\log q}{\epsilon^2}\right)\,.
\end{equation}

\paragraph*{Quantum speedup of Monte Carlo sampling}
The quantum algorithm of Ref. \cite{montanaro2015quantum} provides a quadratic improvement in the dependence of the complexity Eq. \ref{eq:classicalscaling} of Monte Carlo sampling on the precision $\epsilon$.
More precisely, let us assume that $0\le Q(k)\le 1$ for any $k=1,\,\ldots,\,N$ (this can always be achieved by a suitable linear redefinition of $Q$).
Let $U$ be a quantum unitary operator that implements a unitary dilation of the classical algorithm for the Monte Carlo sampling of the Markov chain.
We can assume that the complexity of $U$ has at worst a constant overhead with respect to the classical algorithm, and therefore has the same scaling Eq. b\ref{eq:compls}.
Then, Theorem 2.3 of \cite{montanaro2015quantum} implies that, for any $0<\epsilon<1$ and any $0<\delta<1$ there exists a quantum algorithm that, with $O\left(\left.\log\frac{1}{\delta}\right/\epsilon\right)$ applications of $U$, outputs $\mu\in\mathbb{R}$ such that $\left|\mu - \langle Q\rangle\right|<\epsilon$ with probability at least $1-\delta$.
The complexity of the algorithm is therefore
\begin{equation}
O\left(\frac{n (t/h)\log\frac{1}{\delta}\log q}{\epsilon}\right)\,,
\end{equation}
with the promised quadratic improvement in the $\epsilon$-dependence with respect to Eq. \ref{eq:classicalscaling}.

\section{Classical algorithms for Principal Component Analysis}
\label{app:PCA_classical}
Here we show that the singular vectors and singular values of the data matrix $\mathbf{X}$ cannot be efficiently estimated with classical methods whenever $\left\||x_j\rangle\right\|^2$ is exponentially small in the number of nodes for any time step, \textit{i.e.,} if the size of the support of the probability distribution is always exponential.
More precisely, we show that none of the entries of $\mathbf{X}^\dag\mathbf{X}$ can be estimated efficiently.
Indeed, if $x_j(k)$ is the probability of the $k$-th state at the time step $j$, the $jj'$ entry of $\mathbf{X}^\dag\mathbf{X}$ is
\begin{equation}
    \left(\mathbf{X}^\dag\mathbf{X}\right)_{jj'} = \sum_k x_j(k)\,x_{j'}(k)\,,
\end{equation}
and is equal to the probability that, in a couple of independent Monte Carlo simulations of the Markov chain, the state of the first simulation at the step $j$ is equal to the state of the second simulation at the step $j'$.
This probability can be estimated by running many couples of simulations of the Markov chain.
However, the estimate will be zero until a couple with the state of the first simulation at the step $j$ equal to the state of the second simulation at the step $j'$ is found.
This event will typically happen after
\begin{equation}
    O\left(\frac{1}{\left(\mathbf{X}^\dag\mathbf{X}\right)_{jj'}}\right) \ge O\left(\frac{1}{\max_i\left\||x_i\rangle\right\|^2}\right)
\end{equation}
runs, which is exponentially large in the number of nodes if $\left\||x_j\rangle\right\|^2$ is exponentially small for any time step.

\section{Additional simulations and figures}
\label{app:simulations}

\subsection{Epidemic simulations of social opinion}
Continuous time Markov chains can also be implemented to simulate the spread of social opinion. Here, we consider a model where nodes can exist in one of three states: conservative, liberal, or undecided. As in our analysis on viral epidemics, we perform simulations on the same 7-node network as in the main text. Similar to before, transitions from undecided to liberal or conservative occur at rates dependent on the strength of connection to other liberal or conservative nodes respectively. The data matrix is analyzed between days one and two of the "social epidemic", where at day zero, all states are equally likely. 

\begin{figure}[ht]
    \centering
    \includegraphics{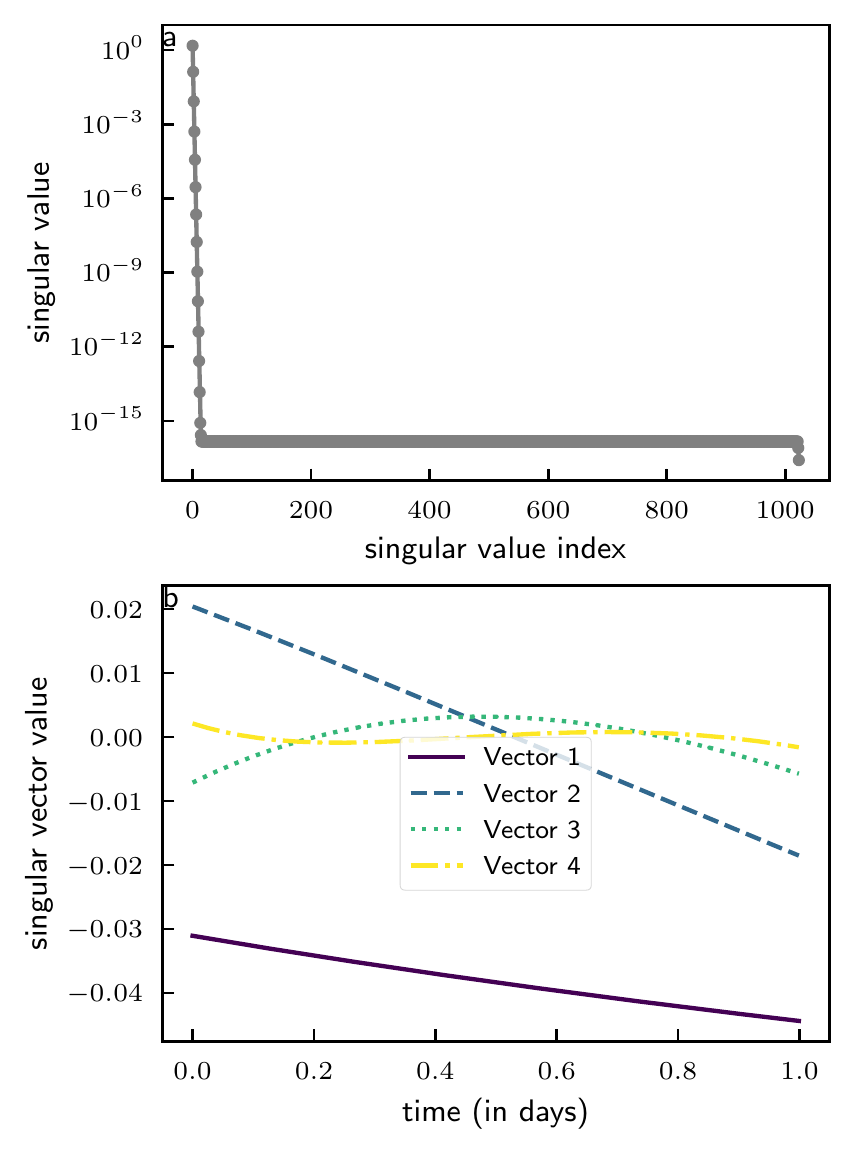}
    \caption{Quantum principal component analysis of the spread of social opinion. \textbf{(a)} Singular values of data matrix $\boldsymbol{X_{mat}}$ decay exponentially fast. \textbf{(b)} Values of the right singular vectors scaled by the square root of their corresponding singular values show the progression of social opinion over time. The results are consistent with prior results for the analysis of viral epidemics on the same network. }
    \label{fig:main_social_singular}
\end{figure}

As shown in Fig. \ref{fig:main_social_singular}a, the data matrix in this case is similarly low rank. Furthermore, we see similar progressions over time in the right singular vectors as shown in Fig. \ref{fig:main_social_singular}b. The first singular vector corresponds to steady state contributions, whereas later singular vectors chart the most prominent changes in the data matrix over time.

\subsection{Model of social distancing}

\begin{figure}[ht]
    \centering
    \includegraphics{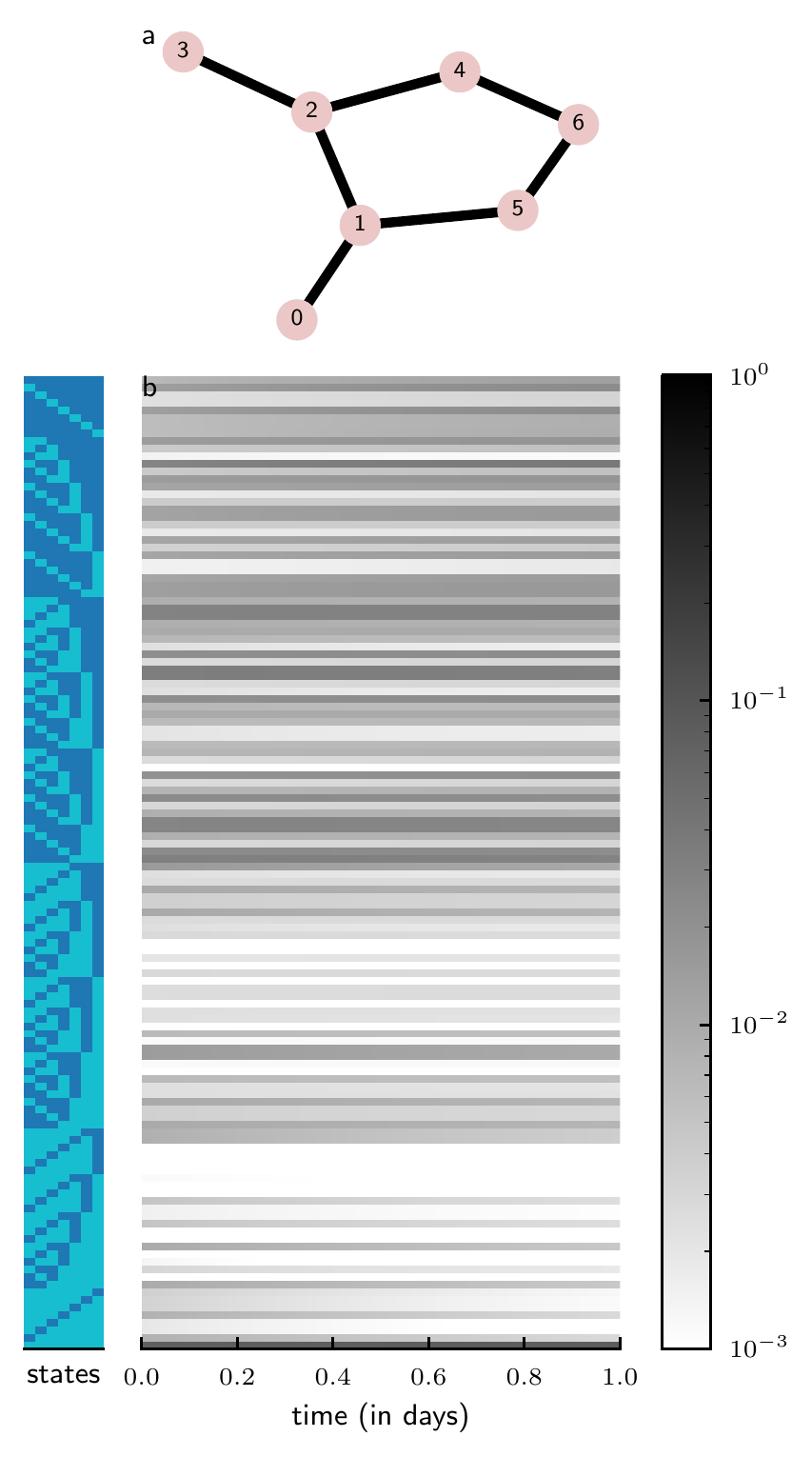}
    \caption{\textbf{(a)} The 7 node network used for simulation of continuous time Markov chain where transitions from susceptible to infected and vice-versa occur with rates $r_{SI}=1.5$, reduced by a factor of five when four or more nodes are infected. \textbf{(b)} Probabilities of Markov states over time shown as colorbar chart (logarithmically scaled). States are enumerated as rows on the left hand side, each denominated by a 7 node colorbar numbered node 0 on the left to node 6 on the right. Dark/light color indicates a node in that state is infected/susceptible respectively. }
    \label{fig:app-distancing}
\end{figure}{}

Supplementary to the main text, we show results here for a simulation of a Markov chain which incorporates effects of "social distancing" in the Markov model. Specifically, we simulate a viral epidemic on the same network as in the main text where transitions from susceptible to infected and vice-versa occur with rates $r_{SI}=1.5$ and $r_{IS}=0.33$ respectively as long as three or fewer nodes are infected. When four or more nodes are infected, "social distancing" is enacted and transitions from susceptible to infected occur at a fifth of the original rate ($r_{SI}=0.3$). As expected, this shifts the steady state away from situations where all nodes are infected to those where four nodes are infected.

The complete progression of this model is plotted in Fig. \ref{fig:app-distancing}b. Note, that the state where all nodes are infected is now unlikely; instead, the states where four or five nodes are infected become very likely (\textit{i.e.,} the social distancing works).

\begin{figure}[ht]
    \centering
    \hspace*{-1.5cm}
    \includegraphics{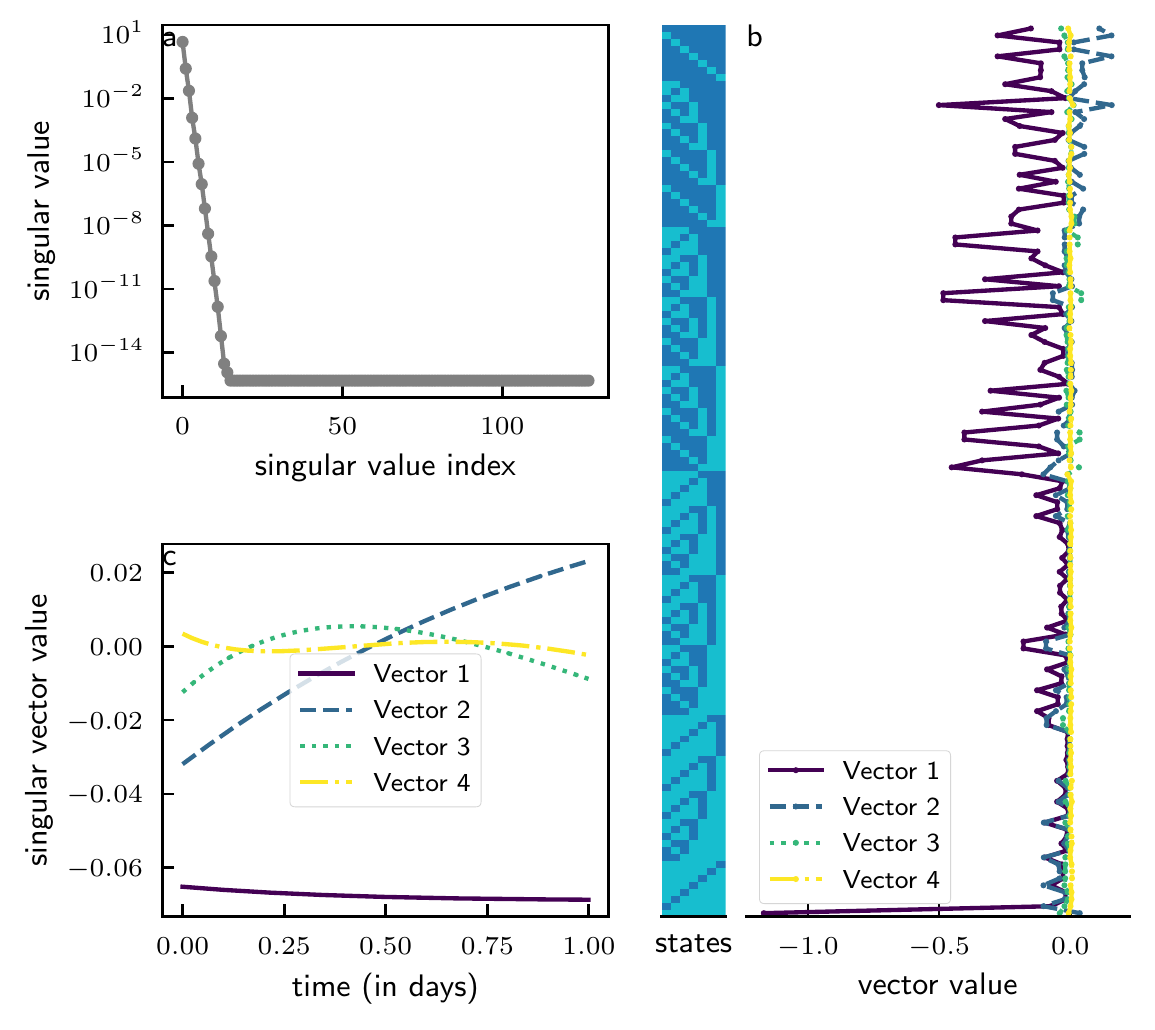}
    \caption{\textbf{(A)} Singular values of data matrix $\boldsymbol{X}$ decay exponentially fast. \textbf{(B)} First four left singular vectors scaled by the square root of their corresponding singular value show that states where four nodes are infected become prominent as this is the inflection point for social distancing. States are enumerated as rows on the left hand side, each denominated by a 7 node colorbar numbered node 0 on the left to node 6 on the right. Dark/light colors indicate a node in that state is infected/susceptible respectively. \textbf{(C)} Values of the right singular vectors scaled by the square root of their corresponding singular value show the progression of the epidemic over time. The first singular vector depicts the steady state and the next few singular vectors detail the intermediate course of the Markov process. }
    \label{fig:app-distancing-singular}
\end{figure}{}

As with the original model, singular values decay exponentially rapidly (see Fig. \ref{fig:app-distancing-singular}a). The first four left singular vectors are plotted in Fig. \ref{fig:app-distancing-singular}b scaled by their corresponding singular value. The steady state singular vector has clearly changed with respect to the original model. Analysis of the first singular vector shows that the dominant states are those where no node is infected and four nodes are infected (\textit{i.e.,} the transition point of social distancing).

\section{Markov states incorporating more than just simple nodes}

\begin{figure}[ht]
    \centering
    \includegraphics{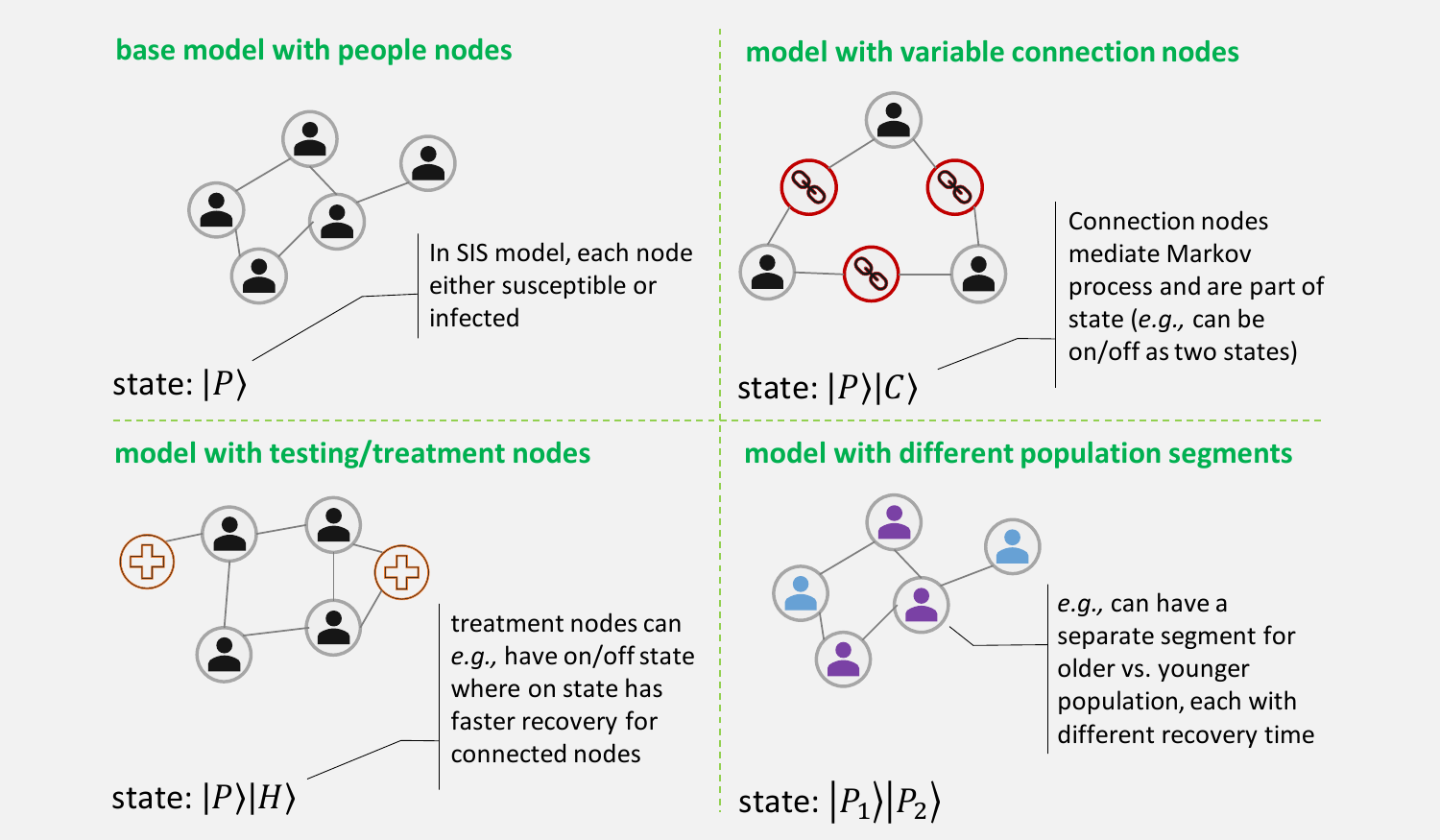}
    \caption{By customizing the properties of nodes in a network, Markov states can model various different phenomenon. Here, we show some of the options available to expand the functionality of a Markov state. In these cases, states are stored in a tensor product structure where information corresponding nodes of different types can be stored in separate registers.}
    \label{fig:node-options}
\end{figure}

Markov models can incorporate nodes of different types which interact in a customized fashion. Fig. \ref{fig:node-options} outlines some of the different options available to one modeling epidemiological processes. Utilizing quantum algorithms, nodes of different types can be stored in separate registers. Analysis and post-processing of the Markov states using a quantum state can take advantage of the structure inherent in these expanded models.

{}

\section{Experimental details}
\label{app:exp_details}
All experiments were performed in Python using the packages Numpy \cite{harris2020array} and Scipy \cite{2020SciPy-NMeth}. For simulations of epidemic spreading, to construct the transition matrix $\boldsymbol{Q}$ for our experiments, we use the method detailed in \cite{simon2011exact}. We assume transitions from infected to susceptible (\textit{i.e.,} recovery rate) occur at rate $r_{IS}=0.33$ indicating that it takes about three days on average to recover from infection. Transitions from susceptible to infected occur at a rate $r_{SI} \in \{ 0.4, 0.8, 1.6 \}$ depending on a node's connection strength to a neighboring node. 

To simplify the exposition, our experiments and simulations were performed classically. In Appendix \ref{app:runtime_proofs}, we further discuss, the asymptotic runtimes and errors for performing our algorithms on a gate based quantum computer. We also show that input states for Markov models can typically be efficiently constructed using simple quantum operations.


\end{appendices}

\end{document}